\documentclass[a4paper,12pt]{article}
\usepackage{amssymb,amsmath}
\usepackage{amsthm}
\usepackage{mathrsfs}
\usepackage{amsfonts}
\usepackage{hyperref}
\usepackage{comment}
\usepackage{fullpage}
\usepackage{enumitem}
\usepackage{color}
\makeatletter
\def\namedlabel#1#2{\begingroup
    #2%
    \def\@currentlabel{#2}%
    \phantomsection\label{#1}\endgroup
}
\makeatother

\newcommand{\Sn}{\mathcal{S}}
\newcommand{\Un}{\mathcal{U}}
\newcommand{\Vn}{\mathcal{V}}

\newcommand{\Jn}{\mathcal{J}}

\newcommand{\F}{\mathcal{F}}

\newcommand{\OA}{\mathrm{OA}}

\newcommand{\eaves}{E}

\newtheorem{definition}{Definition}[section]
\newtheorem{lem}[definition]{Lemma}
\newtheorem{theorem}[definition]{Theorem}
\newtheorem{example}[definition]{Example}
\newtheorem{rem}{Remark}
\newtheorem{corollary}[definition]{Corollary}

\newenvironment{lemma}{\begin{lem} \upshape}{\end{lem}}
\newenvironment{defn}{\begin{definition} \upshape}{\end{definition}}
\newenvironment{thm}{\begin{theorem} \upshape}{\end{theorem}}
\newenvironment{cor}{\begin{corollary} \upshape}{\end{corollary}}

\begin{document}
	
\title{A Combinatorial Model of Interference in Frequency Hopping Schemes}

\author{Mwawi M. Nyirenda\footnote{Mwawi M. Nyirendda was funded by the 
Schlumberger foundation, Faculty for the future scholarship.} \and Siaw-Lynn Ng \and Keith M. Martin 
}

\date{\today}

\maketitle

\begin{abstract}
In a frequency hopping (FH) scheme users communicate simultaneously
using FH sequences defined on the same set of frequency channels. An
FH sequence specifies the frequency channel to be used as
communication progresses. Much of the research on the performance of
FH schemes is based on either pairwise mutual interference or
adversarial interference but not both. In this paper, we evaluate the
performance of an FH scheme with respect to both \textit{group-wise
  mutual interference} and \textit{adversarial interference}
(jamming), bearing in mind that more than two users may be
transmitting simultaneously in the presence of a jammer. We
establish a correspondence between a cover-free code and an FH
scheme. This gives a lower bound on
the transmission capacity. Furthermore, we specify a jammer model and
consider what additional properties a cover-free code should have to
resist the jammer. We demonstrate that a purely combinatorial approach is
inadequate against such a jammer, but that with the use of
pseudorandomness, we can have a system that has high throughput as
well as security against jamming.
\end{abstract}

\section{Introduction}
Frequency hopping is a modulation technique that employs frequency
hopping (FH) sequences in spread spectrum transmission. This
technology was first introduced to allow multiple users to be
co-located within the same spectrum. It also mitigates against
interference from unauthorized users, on the assumption that the
unauthorised users have no knowledge of the FH sequences being used
\cite{Proakis95}. Frequency hopping is widely used in signal
transmission such as Wi-Fi, Bluetooth and ultrawideband (UWB)
communications \cite{Lansford01,Fan96,Sarwate94}. FH sequences specify
the frequency channels on which a transmitter/receiver sends/receives
data as transmission progresses. The main requirement for the
transmitter-receiver pair to communicate is that they need to be on
the same frequency channel at the same time, that is, a communicating pair
of users need to share an FH sequence. When a number of users employ
FH sequences which are defined on the same set of frequency channels,
they form a \textit{frequency hopping multiple access} (FHMA)
system. We will consider properties of FH sequences for FHMA systems
when used in the presence of adversarial interference.
        
We will introduce notation and some background as well as related work 
before describing our contribution.

\section{Background}
\subsection{Frequency hopping schemes}

Let $ \F = \{ f_0, f_1, \ldots, f_{m-1} \} $ be a set of
$m$ frequency channels available to an FHMA system.  We call $\F$ a
\textit{frequency library}. 
        
\begin{defn}
A \textbf{frequency hopping $ (\mbox{FH})$ sequence} of length $v$ over 
$\F$ is a sequence $ X
= ( x_t )_{ t = 0 }^{ v - 1 }$, $x_t \in \F$, $t=0, \ldots, v-1$. 
\end{defn}

We write $X = (x_t)$ if there is no ambiguity.
        
\begin{defn}
A $(v, k, m)$-\textbf{frequency hopping scheme} $ ( (v, k, m
)\mbox{-FHS}) $, is a set $\Sn = \{ X_i : 0 \leq i \leq k - 1\}$ of
size $ k $ where $X_i$ is an FH sequence of length $v$ over a
frequency library $ \F $ of size $ m $.
\end{defn}

Let $ \Sn $ be a $ (v, k, m ) $-FHS. A transmitter and a receiver
share an FH sequence $ X = ( x_t ) \in \Sn $. The channel to be used
for transmission/reception at each time slot $t$ is given by $ x_t $.

\subsection{Pairwise mutual interference} \label{sub:mutual}
The use of the same frequency channel at
the same time by two FH sequences (or more) causes
\textit{interference}. \textit{Pairwise mutual interference} is where
two FH sequences in an FH scheme interfere with each other. It is
measured by Hamming correlation. Formally, let $ \Sn $ be a $ (v, k, m
)\mbox{-FHS}$ and let $ X, Y \in \Sn $, $X=\left(
x_0,x_1,\ldots,x_{v-1}\right) $ and $Y=\left(
y_0,y_1,\ldots,y_{v-1}\right) $. The \textit{Hamming correlation} $
H_{XY} $ at \textit{relative time delay $ \tau $} between $ X $ and $
Y $ is

        \begin{equation}
                H_{XY}(\tau)=\sum_{i=0}^{v-1}h(x_i,y_{i+\tau}),\;\;\; 0 \leq \tau < v, 
        \end{equation}  
        where 
        \[ h(x,y) = \left\{
                \begin{array}{l l}
                        1 & \quad \mbox{ if } x = y,\\
                        0 & \quad \mbox{ if } x\neq y.
                \end{array} \right.\]

The operations on indices are performed modulo $v$. When $ X
= Y $ we write $ H_X(\tau) $ for $ H_{XX}(\tau) $ and this is referred to as the
\textit{Hamming auto-correlation} of $ X $.  If $X \neq Y$ we call 
$H_{XY}(\tau)$ the \textit{Hamming cross-correlation}.
        
        We define the \textit{maximum out-of-phase Hamming auto-correlation} of an FH sequence $ X \in \Sn $ as
        \begin{equation*} 
                H(X)=\max_{1 \leq \tau < v}\{H_{XX}(\tau)\}, 
        \end{equation*}
        and the \textit{maximum Hamming cross-correlation} between any two distinct FH sequences $ X $ and $ Y $ in $ \Sn $ as
        \begin{equation*} 
                H(X,Y) = \max_{ 0 \leq \tau < v } \{ H_{XY}( \tau ) \}.
        \end{equation*}
        Further, we define
        \begin{equation*} 
                M(X,Y)=\max \{H(X),H(Y),H(X,Y)\}. 
        \end{equation*}

Lempel and Greenberger ~\cite{lempel74} developed the following bound on
the maximum out-of-phase auto-correlation of a sequence:
        
        \begin{lemma}\cite[Lemma 4]{lempel74} \label{Lempellemma1} For every sequence $ X = (x_t) $ of length $v$ over $\F$, $|\F| = m$, 
                \begin{equation}\label{Lempelbound1}
                        H(X)\ge \frac{(v-r)(v+r-m)}{m(v-1)}
                \end{equation}
                where $ r \equiv v \mod m $.
        \end{lemma}
        
For a $(v,k,m)$-FHS, $\Sn$, the \textit{maximum periodic Hamming auto-correlation of $ \Sn $}, $ H_a(\Sn) $, and the \textit{maximum periodic Hamming cross-correlation of  $ \Sn $}, $ H_c(\Sn) $, are defined as
        \begin{equation*}
                H_a(\Sn) = \max \{ H(X)| X \in \Sn \},
        \end{equation*} 
        \begin{equation*}
                H_c(\Sn) = \max \{ H(X, Y)| X, Y \in \Sn, X \ne Y \}.
        \end{equation*} 
        The \textit{maximum Hamming correlation of $ \Sn $} is defined as
        \begin{equation}\label{mHc}
                H_m(\Sn) = \max \{ H_a(\Sn), H_c(\Sn) \}.
        \end{equation} 
      Peng and Fan~\cite{Peng04} gave the following bound for the maximum Hamming correlation of a set of sequences:
        \begin{lemma}\cite[Corollary 1]{Peng04} \label{Pengfanlemma1}
                Let $\Sn$ be a $(v, k, m)$-FHS. Let $ I = \left\lfloor \frac{vk}{m} \right\rfloor $. Then
                \begin{equation}\label{PengFanBound2}
                        H_m(\Sn) \ge \left\lceil\frac{2Ivk-(I+1)Im}{(vk-1)k} \right\rceil.
                \end{equation}
        \end{lemma}

An FH sequence $ X$ is said to be optimal in the
Lempel-Greenberger bound if the bound (\ref{Lempelbound1}) is met, and
a $ (v, k, m ) $-FHS is said to be an optimal FH scheme in the
Peng-Fan bound if the bound (\ref{PengFanBound2}) is met. 

There are many FH sequence constructions that are optimal in the
Lempel-Greenberger bound or the Peng-Fan bound. We list a few here
that use techniques from algebra, combinatorial designs as well as codes. 
Lempel and
Greenberger \cite{lempel74} construct optimal FH sequences
from algebraic transforms of m-sequences. Fuji-Hara et
al. \cite{FujiHara04} provide constructions of optimal sequences using affine
geometries, cyclic designs and difference families. Using cyclotomy as well as
quadratic residues, Chung and Yang \cite{Chung10} obtain optimal FH
sequences with new parameters. 
Ding et al. \cite{Ding09} employ Reed-Solomon codes to obtain
sets of FH sequences optimal in the Peng-Fan bound. Using cyclotomy and the
Chinese remainder theorem, Ren et al. \cite{Ren14} also obtain sets of FH
sequences meeting the Peng-Fan bound. Ren's constructions are a
generalisation of the constructions in \cite{Ding09} and
\cite{Zhang09} which also use cyclotomy over finite fields. Ding et
al. \cite{Ding09} also use the trace function to construct sets of FH
sequences that are optimal in the Peng-Fan bound. The FH sequences
constructions given above are all optimal in the sense of
meeting either the Lempel-Greenberger bound or the Pen-Fan
bound. However, both bounds are based on \textit{pairwise} Hamming
correlation. We will explain in Section \ref{Contribution} the
insufficiency of using pairwise correlation and propose 
using Hamming \emph{group} correlation to measure \textit{group-wise}
mutual interference.

\subsection{Adversarial interference}
        
Interference originating from unauthorised entities where signals are
deliberately transmitted to interfere with legitimate transmission is
called \textit{adversarial interference} or \textit{jamming}. We discuss this in more detail in Section \ref{attacker}.
        
Now we describe the work of Bag et al. \cite{Bag13}, Nyirenda et
al. \cite{Nyirenda14} and Emek and Wattenhofer \cite{Emek13} who
focused on adversarial interference rather than mutual
interference. What is common across these three constructions is the
use of pseudorandomness.  However, the capabilities of the jammers
differ. We will elaborate a little on this.
 
Bag et al. \cite{Bag13} use mutually orthogonal Latin squares to
obtain sets of FH sequences. These FH sequences achieve maximum
transmission rate of 100\% without adversarial interference. A 
transmitter and a receiver share
a pair of secret pseudorandom numbers before the start
of communication and uses them for the entire session. The jammer is
assumed to be able to jam at most a certain number of the frequency
channels.  However, it was shown in \cite{Nyirenda14} that a
jammer only needs to eavesdrop on a single time slot to obtain the
pair of secret shared pseudorandom numbers. This allows the jammer
to derive the FH sequences and thus interfere with any FH sequence of
its choice.  This weakness was amended in \cite{Nyirenda14}, where it
was proposed that there should be a new secret pseudorandom number for
each time slot.  We will review one of the schemes of
\cite{Nyirenda14} in Section \ref{Discuss} in light of our model described in
Section \ref{System1}.

Emek and Wattenhofer \cite{Emek13} construct FH sequences as a random
walk on an expander graph. The authors consider a single pairwise
communication where subsequent channels for transmission are included
in the data transmitted. A jammer can eavesdrop and jam a certain
fraction of the available frequency channels. In this paper two
adversarial models were considered. In the first model, a jammer can
only acquire information about the channel (but not the content) that
was used in previous time slots after a certain number of time slots
have lapsed, while in the second model it has knowledge of both
channel and content. Knowing the transmitted messages is important
since the content specifies the next channels. At any time slot, the
FH sequence is guaranteed successful transmission with some minimum
probability.  However, it is not clear what happens if more than one
pair (transmitter/receiver) of communication occurs simultaneously.

\subsection{Our contributions}\label{Contribution}
        
In this paper, we evaluate the performance of an FH scheme with
respect to both group-wise mutual interference \textit{and}
adversarial interference. This framework was introduced in
\cite{Nyirenda14}, motivated by the fact that more than two pairs of 
users may be
transmitting simultaneously in the presence of a jammer.
This means that measuring pairwise mutual interference, while
giving some idea of the throughput of the scheme, is not adequate.
   
An overview of our contributions is as follows. We refine the system
and jammer model which was introduced in \cite{Nyirenda14}. We
establish a correspondence between a cover-free code and an FH
scheme. We show that when a cover-free code is considered as an FH
scheme, a user can successfully transmit in at least a specified
fraction of time in the presence of a given number of interfering FH
sequences. We specify a jammer model for an FH scheme. Considering the
resources and knowledge of a jammer, we look at how an FH scheme can
mitigate against the jammer. We examine necessary and desirable
additional properties of cover-free codes such that they can be used
in the presence of adversarial interference: a cover-free code allows
us to determine the throughput of an FH sequence in the presence of
other interfering FH sequences but not in the presence of adversarial
interference. Finally we discuss the limitations of cover-free codes
against a jammer, and demonstrate that one effective way to improve
resistance against a jammer is to use pseudorandomness.

The rest of the paper is organised as follows. In Section
\ref{System1} we introduce the system and jammer model and the necessary
notation. In Section \ref{CFCFHS} we introduce cover-free codes and
show their equivalence to FH schemes. Section \ref{CFCFHS} also
examines the additional properties of cover-free codes to defend
against jamming. In Section \ref{Discuss} in particular we discuss how
pseudorandomness can be used to strengthen cover-free codes to
withstand a jammer so that the FH sequences can be used for longer
periods of time. We conclude in Section \ref{Conclude}.

\section{System and jammer model}\label{System1}
        
\subsection{System model}\label{System}
        
Let $ \Sn $ be a $ (v, k, m ) $-FHS. A single FH sequence is used by a
single transmitter/receiver pair to communicate in the presence of
both mutual and adversarial interference. We first consider the case
of mutual interference.
        
\begin{defn}\label{defn:hgc}
The \textbf{Hamming group correlation} $ G ( X, \Un) $ between an FH
sequence $ X \in \Sn $ and the FH sequences in $ \Un \subseteq \Sn
\setminus \{ X \} $, $ | \Un | = w $, $ 1 \leq w < k $, is defined as
the number of time slots in $ X $ that use the same frequency channels
as the corresponding time slots of some FH sequence in $ \Un $:
                \begin{equation}\label{GD}
                G ( X, \Un ) = \left| \{ x_t | \exists Y \in \Un \mbox{ s.t. } x_t = y_t, t = 0, \ldots, v - 1 \} \right|.
                \end{equation}
        \end{defn}

Note that the Hamming group correlation is a generalisation of
Hamming correlation. When $ | \Un | = 1 $, say $\Un = \{Y\}$, 
then $ G( X, \Un )= H_{X Y} (0) $.  Note also that in the following, while
the definitions and results work for $w=0$, we will assume that $w > 0$
for the idea of group correlation to be meaningful.

\begin{lemma} \label{lemma:hgcbound}
Let $\Sn $, $\Un$, $X$ be as above.  Let $H_m(\Sn)$ be the 
maximum Hamming correlation of $\Sn$.  Then 
$G ( X, \Un ) \le w H_m(\Sn)$.
\end{lemma}
\begin{proof}
Let $\Un = \{Y_1, \ldots, Y_w\}$.  Then
$G ( X, \Un ) \le H_{XY_1}(0) + \cdots +  H_{XY_w}(0)$.  Since
$H_{XY_i}(0) \le H_m(\Sn)$ for $1 \le i \le w$, we have $G ( X, \Un ) \le w H_m(\Sn)$.
\end{proof}

The notion of Hamming group correlation is the complement of the group
distance defined in \cite{Jin07} when a $ (v, k, m) $-FHS is
considered as a set of $ k $ codewords of length $ v $ over $ \F
$. Hamming group correlation $ G ( X, \Un) $ gives the number of
time slots of an FH sequence $ X $ that are blocked by the FH
sequences in the $ w $-subset $ \Un $ of $ \Sn $.
        
We define a \textit{session} as being made up of $ v $ time slots,
that is one full length of an FH sequence. We now define the
throughput of an FH sequence. 
        
\begin{defn}\label{th1}
Let $ \Sn $ be a $ ( v, m, k ) $-FHS. Let $X \in \Sn$ and let
$ \Un \subseteq \Sn \setminus \{X\}$, $|\Un | = w$, $1 \leq w < k $. 
Then the
$ w $-\textbf{throughput of an FH sequence} $ X $ with respect to $
\Un $ is the rate of successful transmission of $ X $ in a session in
the presence of FH sequences in $ \Un $:
        \begin{equation}\label{TH} 
                \rho_w ( X, \Un ) = 1 - \frac{ G ( X, \Un ) }{ v }. 
        \end{equation}
\end{defn}

It is desirable that $\rho_w ( X, \Un )$ be large, which means an FH
sequence transmits successfully in many time slots.  Here we focus on the
worst-case $w$-throughput of a $(v,k,m)$-FHS, which gives the lowest
possible throughput over a session in a communication system. We first
look at the case without a jammer. The worst-case $w$-throughput
in the presence of a jammer will be dealt with in Section
\ref{attacker}.
        
\begin{defn}
        Given a $ (v, k, m) $-FHS, $ \Sn $, \textbf{the worst-case $ w $-throughput of an FH sequence} $ X \in \Sn $ is
        \begin{equation}\label{wct}
                \hat{\rho}_w(X, \Sn) = \min\limits_{\stackrel{\Un \subseteq \Sn \setminus \{X\}}{|\Un|=w}} \left\{  \rho_w(X,\Un) \right\} .
        \end{equation}
\end{defn}

So, given an FH sequence $ X \in \Sn $, (\ref{wct}) gives the minimum
number of time slots in which the FH sequence can transmit data if
some other $ w $ FH sequences in $ \Sn $ are also in use.
        
Next we consider the worst-case $w$-throughput of a particular subset
of a $(v, k, m)$-FHS.

\begin{defn}
Let $\Sn$ be a $(v, k, m) $-FHS and let $\Vn \subseteq \Sn $, $| \Vn | = w+1$.
        The \textbf{worst-case $w$-throughput of $\Vn$} is
        the minimum number of time slots in which any $ X
        \in \Vn $ can transmit information if the other $ w $ FH sequences in $
        \Vn \setminus\{X\} $ are in use:
        \begin{equation}
                \hat{\rho}_w(\Vn) = \min\limits_{X \in \Vn} \left\{  \rho_w(X, \Vn \setminus \{
 X \})\right\} .
        \end{equation}
\end{defn}

We conclude with the worst-case $ w $-throughput of an FHS.

\begin{defn}
        The \textbf{worst-case $ w $-throughput of a $(v, k, m)$-FHS,
          $ \Sn $}, is the minimum of the values $ \hat{\rho}_w(\Vn) $
        for each $(w + 1)$-set $ \Vn \subseteq \Sn $:
        \begin{equation}\label{worsts}
                \hat{ \rho }_w ( \Sn ) = \min\limits_{ \stackrel{\Vn \subseteq \Sn}{ | \Vn | 
 = w + 1} } \left\{ \hat{\rho}_w(\Vn) \right\}.
        \end{equation}
\end{defn}

We write $ ( v, k, m; \hat{ \rho }_w ( \Sn ) ) $-FHS for a $ ( v, k, m
) $-FHS, $ \Sn $, with worst-case $ w $-throughput $ \hat{ \rho }_w (
\Sn ) $. For any $ X \in \Sn $ we can estimate the worst-case $ w
$-throughput of $ X $:

\begin{lemma} \label{lemma:lower1}
Let $\Sn$ be a  $ (v, k, m) $-FHS and let $X \in \Sn$.  Then
$ \hat{\rho}_w (X, \Sn )  \ge \hat{\rho}_w ( \Sn )$.
\end{lemma}
\begin{proof}
Suppose $ \hat{\rho}_w (X, \Sn )  < \hat{\rho}_w ( \Sn )$, that is,
there is some $(w+1)$-subset $\Vn \subseteq \Sn$ containing $X$ with
$\rho_w(X, \Vn \setminus \{X\}) < \hat{\rho}_w ( \Sn )$.  This contradicts
the definition of $\hat{\rho_w}(\Sn)$.
\end{proof}

Note however, that $\hat{\rho}_w(\Vn)$ and $\hat{\rho}_w(X, \Sn)$ are not
necessarily comparable.  For example, suppose a $(w+1)$-subset $\Vn$
containing $X$ satisfies 
$\hat{\rho}_w(X, \Sn) = \rho_w(X, \Vn \setminus \{X\})$.  This does not
rule out the existence of an FH sequence $Y \in \Vn$ such that 
$\hat{\rho}_w(\Vn) = \rho_w(Y, \Vn \setminus \{Y\}) < 
\rho_w(X, \Vn \setminus \{X\})$.  Similarly, suppose an FH sequence $X$
satisfies $\hat{\rho}_w(\Vn) = \rho_w(X, \Vn \setminus \{X\})$.
This does not rule out the existence of another $(w+1)$-subset 
$\Vn' \subseteq \Sn$ with 
$\hat{\rho}_w (X, \Sn) = \rho_w(X, \Vn' \setminus \{X\}) < 
\rho_w(X, \Vn \setminus \{X\})$.

We can also relate the worst-case $ w $-throughput of an FH
scheme with the maximum Hamming correlation of an FH scheme. Given a $
(v, m, k) $-FHS, $ \Sn $ with  maximum Hamming correlation $H_m ( \Sn )$, 
we can obtain a lower bound on the worst-case $ w $-throughput of $ \Sn $:

\begin{lemma}
        \begin{equation}\label{wcHc1}
                \hat{\rho}_w ( \Sn ) \ge 1 - \frac{ w \cdot H_m ( \Sn ) }{ v }.
        \end{equation}
\end{lemma}
\begin{proof}
By definition, $ \hat{\rho}_w ( \Sn ) = \rho_w(X, \Vn \setminus \{X\})$
for some $(w+1)$-subset $\Vn \subseteq \Sn$ containing $X$.  The 
inequality follows from Definition \ref{defn:hgc} and Lemma
\ref{lemma:hgcbound}.
\end{proof}

It also follows from Lemma \ref{lemma:lower1} that
$\hat{\rho}(X, \Sn)  \ge 1 - \frac{ w \cdot H_m ( \Sn ) }{ v }$.

\subsection{Jammer model}\label{attacker}
We consider the presence of a jammer who sends noisy signals
on frequency channels to block the signal transmissions of legitimate
users. The jammer knows $ \F $, the $ ( v, k, m ) $-FHS, $ \Sn$, and
the number of FH sequences used in a session, $ w + 1$ $( 0 \le w < k)
$. However, the jammer has no knowledge of the actual FH sequences to
be used. Its strategy is to eavesdrop and jam. At each time slot it
has enough resources to eavesdrop on $\theta_1 m$ channels, $ 0 \leq
\theta_1 \leq 1 $, and jam on $\theta_2 m$ channels, $ 0 \leq \theta_2
< 1 $. (We assume that it cannot jam all the frequency channels at each
time slot.) It can use the information it acquires while eavesdropping
to make choices of which channels to jam. This jammer is described
as a \textit{$(\theta_1, \theta_2)$-adaptive jammer}. When a signal is
jammed, legitimate users hear noise and acknowledge failure of
transmission. So we treat a jamming signal as an erasure.  The goal of
the jammer is to reduce the worst-case $ w $-throughput of $\Sn$.

(We note here that we have allowed the possibility of $w=0$ here, that
is, only one FH sequence is used.  
While it is not meaningful to talk
about correlation when there is only one sequence, 
it is still perfectly reasonable for a jammer to 
want to identify that one sequence.)

We model a jammer's channel selection for jamming as a set of FH
sequences $\Jn = \{Y_i| i=0, \ldots, \theta_2 m -1\}$, where $Y_i$ is
an FH sequence of length $v$ over $\F$. We will call $\Jn$ the
\textit{jamming sequences}.
        
\begin{defn} \label{defn:jammerTP}
Let $\Sn$ be a $(v,k,m)$-FHS over $\F$.  Let $X \in \Sn$ and  
let $\Un \subseteq \Sn \setminus \{X\}$, $|\Un|=w$, $ 0 \le w < k$.
Suppose $\Jn$ is a set of $\theta_2m$ jamming sequences of length
$v$ over $\F$.  Then the \textbf{$ ( w, \Jn ) $-throughput} of $ X $ in the presence of \textit{both} legitimate FH sequences in $ \Un$ and jamming sequences $ \Jn $ is:
        \begin{equation}\label{THJ} 
                \rho_{ w, \Jn } ( X, \Un) = 1 - \frac{ G ( X,  \Un \cup \Jn  ) }{ v }. 
        \end{equation}
\end{defn}

The rest of the measures introduced in Section \ref{System} can be
easily modified to measure the worst case throughput of a $(v,
k,m)$-FHS in the presence of \textit{both} mutual interference and
jamming. These are summarised in Table \ref{table:JammerTHR}.

\begin{table}[h]
        \centering 
        \begin{tabular}{|p{0.4\textwidth}|p{0.5\textwidth}|}\hline
                worst-case $(w,\Jn)$-throughput of $ X \in \Sn$ & $ \hat{ \rho }_{ w, \Jn }( X, \Sn) = \min\limits_{ \stackrel{\Un \subseteq \Sn\setminus\{X\}}{| \Un|=w }} \left\lbrace \rho_{ w, \Jn }( X, \Un ) \right\rbrace $ \\\hline
                worst-case $(w,\Jn)$-throughput of $\Vn \subseteq \Sn$,
$|\Vn|=w+1$ & $ \hat{ \rho }_{ w, \Jn }( \Vn ) = \min\limits_{X \in \Vn} \left\lbrace \rho_{ w, \Jn }( X, \Vn\setminus\{X\}  ) \right\rbrace $ \\\hline
                worst-case $(w,\Jn)$-throughput of $\Sn$ & $ \hat{ \rho }_{ w, \Jn } ( \Sn ) = \min\limits_{\stackrel{\Vn \subseteq \Sn}{ | \Vn |  = w + 1} } \left\lbrace \hat{\rho}_{ w, \Jn } ( \Vn )\right\rbrace  $ \\\hline
        \end{tabular}
        \caption{Performance measures of FH scheme in the presence of mutual interference and jamming. }
        \label{table:JammerTHR}
\end{table}

Suppose $X \in \Sn$ is an FH sequence in use.  Clearly if $X \in \Jn$
then the jammer would have succeeded in reducing the worst case
throughtput to 0.  In Section \ref{jptycfc} we will discuss the
strategy the jammer might adopt to discover $X$.

In the literature, jammers are classified according to their
capabilities (broadband or narrowband) and their behaviour (constant,
random or reactive) \cite{Pelechrinis11,Poisel04,Xu05}. Our
$(\theta_1, \theta_2)$-adaptive jammer includes these jammers. For
example, a broadband jammer means the jammer jams on contiguous set of
channels and so we can consider this as $ \theta_2 m > 1 $
and the jamming sequences $Y_1$, $Y_2$, $\ldots$ have neighbouring channels
in each time slot. If we
consider a constant jammer who always jams on the same channel(s) $f_{i_1}$, 
\ldots, $f_{i_{m\theta_2 }}$, 
then we have a $ (\theta_1, \theta_2) $-adaptive jammer
with $ \Jn=\{Y_j =( f_{i_j},f_{i_j}, \ldots, f_{i_j}) | j=1, \ldots, m\theta_2\} $.

\section[]{Cover-free codes and frequency hopping schemes}\label{CFCFHS}
        
We now model a $ ( v, k, m ) $-FHS as a code and consider the desirable
correlation properties in this light.

Recall that $ \F = \{ f_0, f_1, \ldots, f_{m-1} \} $ is the alphabet over
which we defined the FH sequences.  Let $\Sn$ be a  $ ( v, k, m ) $-FHS 
defined over $\F$.  We may treat each sequence of $\Sn$ as a $v$-tuple
in $\F^v$ and therefore treat $\Sn$ as a code of length $v$ and size $k$
over $\F$.  The Hamming distance $d_H(X,Y)$ between two codewords 
$X$, $Y$, is the number of 
places where the two codewords differ. The following is easy to verify:

\begin{lemma}
Let  $\Sn$ be a $( v, k, m ) $-FHS defined over $\F$ considered
as a code, and let $X, Y \in \Sn$.  Then $d_H(X,Y) = v - H_{XY}(0)$.
The minimum distance $d_{\Sn}$ of $\Sn$ as a code satisfies
$$
 d_{\Sn} = \min_{X,Y \in \Sn} \{d_H(X,Y)\} 
   \ge v - H(X,Y).
$$
\end{lemma}

We will call $\Sn$ a $( v, k, m; \hat{\rho}_w(\Sn)) $-FHS or 
a $(v,k,m; d_{\Sn})$-code (omitting $\hat{\rho}_w(\Sn)$and $d_{\Sn}$ if
they are not known) depending on context.  Clearly
a $( v, k, m ) $-FHS can always be treated as a $ ( v, k, m ) $-code and
vice versa.

The notion of cover-free codes has been used in \cite{Jin07,Kumar99,Stinson01} for blacklisting and traitor tracing schemes. In this paper we use the definition of Staddon et al \cite{Stinson01}.
        
\begin{defn}[Cover-free codes\cite{Stinson01}]\label{defCFC1} 
Let $\Sn $ be a $ ( v, k, m ) $-code. For any subset $ \Sn'
\subseteq \Sn $ and any $ X \in \F^v $, define:
        \begin{equation}\label{CFCs}
                \textit{ I }( X,\Sn') = \{ i : x_i = y_i \mbox{ for some } Y \in \Sn' \}.
        \end{equation}
Let $w \ge 1$ be an integer and let $0 \le \alpha < 1$.  
Then $ \Sn $ is called $ ( w, \alpha ) $-\textbf{cover-free code}, 
denoted $(w, \alpha ) $-CFC, if $ | \textit{I} ( Z, \Sn')|<(1-\alpha)v$ 
for any $ \Sn' \subseteq \Sn $, $ | \Sn' | = w $ and any 
$ Z \in \Sn \setminus \Sn' $.
\end{defn}

We see that $|I(X, \Sn')|$ gives us the number of ``incidents''
between $X$ and the codewords in $\Sn'$.  If we treat $\Sn$ as a $ (
v, k, m ) $-FHS, then for $\Sn' \subseteq \Sn$, $X \in \Sn \setminus
\Sn'$, $|I(X, \Sn')|$ is precisely $G(X, \Sn')$ from Definition
\ref{defn:hgc}. Hence we have a direct correspondence between an FH
scheme with a given Hamming group correlation and a cover-free code.
\begin{thm}\label{correspondence}
Let $\Sn$ be a $ ( v, k, m ) $-code over $ \F $, $ | \F | = m $. Then $ \Sn $ is a $ ( w , \alpha ) $-CFC if and only if $ \Sn $ is a $ ( v, k, m )$-FHS with worst-case $ w $-throughput $\hat{\rho}_w(\Sn)$ greater than $ \alpha $.
\end{thm}

\begin{proof}
Suppose $\Sn$ is a $ ( w , \alpha ) $-CFC.  Then, $|I(X, \Sn')| <(1-\alpha)v$
for all $\Sn' \subseteq \Sn$, $|\Sn'| = w$, $X \not \in \Sn'$.
Since $G(X, \Sn') =|I(X, \Sn')|$, we have 
$\rho(X, \Sn') = 1 - G(X, \Sn')/v > \alpha$ for all 
$\Sn' \subseteq \Sn$, $|\Sn| = w$, $X \not \in \Sn'$.
Since  $\hat{\rho}_w(\Sn) = \rho(X, \Sn')$ for some $X$, $\Sn'$ by definition,
we have  $\hat{\rho}_w(\Sn)>\alpha$.

Conversely, let $\hat{\rho}_w(\Sn)>\alpha$. Suppose $\Sn$ is not a
$ ( w , \alpha ) $-CFC, that is, there exist some $Z, \Sn'$, 
$\Sn' \subseteq \Sn$, $|\Sn| = w$, $Z \not \in \Sn'$, such that
$|I(X, \Sn')| \ge(1-\alpha)v$.  This implies that 
$\rho(Z, \Sn') = 1 - G(Z, \Sn')/v \le \alpha$.  However, by assumption,
$ \hat{\rho}_w(\Sn)>\alpha \ge \rho(Z, \Sn')$, which contradicts
Lemma \ref{lemma:lower1}.  
\end{proof}

Hence the problem of designing frequency hopping schemes with high throughput
is equivalent to finding cover-free codes with low ``incidents''.
Note that a $ ( v, k, m; d) $-code is a $ ( 1, d/v ) $-CFC. We are
interested in the cases where $ w > 1 $.    
It was shown in \cite{Stinson01} that codes with large
minimum distance are cover-free codes.
\begin{thm}\cite[Theorem 4.3]{Stinson01}\label{findw1}
        Suppose that $\mathcal{S}$ is a ($v,k,m;d$)-code such that $d > v ( 1 - \frac{1}{ w^2})$. Then $\mathcal{S}$ is a $( w, 1- \frac{ 1 }{ w } ) $-CFC.
\end{thm}
We have the following as a corollary of Theorems \ref{correspondence} and \ref{findw1}.
\begin{cor}\label{shortresult}
        A $ ( v, k, m; d ) $-code with $ d > v ( 1 - \frac{ 1 }{ w^2 } ) $ 
gives a $ ( v, k, m) $-FHS with worst-case $ w $-throughput greater than 
$ 1 - 1/w $.
\end{cor}

\begin{example} \label{eg:RScodes}
        Let $ v $ and $ w $ be integers where $ v \ge 2 $ and $ w \ge 2 $. Let $ m $ be a prime power such that $m \ge v $. Let $\F $ be the finite field of cardinality $ m $ and let $ \alpha_1, \alpha_2, \ldots, \alpha_v \in \F $ be distinct. Define a length $ v $ Reed-Solomon code $ \Sn $ over $ \F $ by:
                $$ \Sn = \left\lbrace \left( f(\alpha_1), f(\alpha_2), \ldots , f(\alpha_v)\right) : f \in \F[X] \mbox{ and deg }f < \left\lceil 
                \frac{v}{w} \right\rceil \right\rbrace .$$
        Then $ \Sn $ is a $ ( w, 1 - 1/w ) $-CFC code, which is a $ ( v, m^{ \left\lceil \frac{v}{w^2} \right\rceil },m)$-FHS with worst case $w$-throughput
greater than $1-1/w$.
\end{example}

\subsection[]{Jamming resistance properties for cover-free codes}\label{jptycfc}

Theorem \ref{correspondence} shows that cover-free codes give FHSs
with a guaranteed minimum throughput without adversarial presence.  We
now consider how a jammer influences the throughput of such FHSs.  We
delve into further properties that cover-free codes should have to
mitigate a $( \theta_1, \theta_2)$-adaptive jammer. For simplicity, we
assume $ \theta_1 m = \theta_2 m = 1 $.

We introduce some terms and notation first.

Let $ \Sn $ be a $ (v, k, m; \hat{ \rho }_w(\Sn)) $-FHS over $ \F =
\{f_0, \ldots, f_{m-1}\}$. In any session of $v$ time slots, there are
$ w + 1$ FH sequences that are in use by legitimate users. We call
these \textit{active} sequences. Let $\Sn'' \subseteq \Sn $ be the set
of active FH sequences. At any time slot $ t $, $ 0 \le t \le v-1 $,
there are at most $ w + 1 $ frequency channels in use,
which we call \textit{active channels}. At any time slot $t$, let the
multiset $ \F_t = \{x^0_t, \ldots, x^{k-1}_t\} $ denote all the
channels that appear in all the FH sequences in $\Sn$ at that time. The
$m$-tuple $ M_t = (a_0, \ldots, a_{m-1}) $ denotes the multiplicity of
channels $f_0, \ldots, f_{m-1}$ at time slot $ t $, where $ a_i = |
\{j : x^j_t = f_i \} | $.

The aim of a jammer is to reduce the worst-case $w$-throughput of the
FHS.  To this end a jammer would aim to identify an active FH sequence
as quickly as possible. It can then reduce the worst-case $ w
$-throughput to 0, or close to 0. We call the jammer's \textit{search
  space} the set of FH sequences $ \Sn^*_t \subseteq \Sn$ which the
jammer needs to look at to identify an active FH sequence at time slot
$ t $. At the beginning of a session, $ t = 0 $, the search space is
the whole FH scheme, $ \Sn_0^* = \Sn $. Let the number of time slots
it takes a jammer to determine an active FH sequence be denoted $
\gamma v $, $ 0 \le \gamma \le 1 $. It is desirable that $\gamma$ be
large.  The aim of a $ (v, k, m;\hat{\rho}_w( \Sn ) ) $-FHS is to make
the jammer's advantage not much better than a random guess.

We now explore the behaviour of a jammer given $\Sn$ and consider what
properties $\Sn$ should have as a defence.  A jammer can trivially
reduce the worst-case $ w $-throughput of $\Sn$ to $0$ if $\Sn'' =
\Sn$, since all sequences are active and the jammer can choose any $ X
\in \Sn $ to jam the entire session.  As a mitigation strategy we
have:

\begin{description}
        \item[\namedlabel{m1}{\textbf{M1}}] Use only a fraction of $ \Sn $, that is $ \Sn'' \subset \Sn $.
\end{description}

If the jammer does not know which sequences or channels are being used
then it will have to guess which frequency channel to eavesdrop on. At
time $t=0$, there are $k$ FH sequences assumed to be equally likely
over $m$ frequency channels, and for each frequency channel $x_i$ there
are $a_i$ FH sequences of that frequency channel. The probability that
frequency channel $x_i$ is active is:
        \begin{equation}\label{Pguess}
                Prob(x_i \mbox{ is active}) = 1 -  {k - a_i \choose w + 1} / { k \choose w + 1}.
        \end{equation}
The probability in (\ref{Pguess}) is maximum 
for a frequency channel $ x_i $ such that $ a_i \geq a_j $ for all $ i \neq j
$. Therefore, if there exists such an $ x_i $, 
then a jammer will choose it and will have a higher chance of 
jamming an active sequence. 

Hence we propose that:
\begin{description}
        \item[\namedlabel{m2}{\textbf{M2}}] A $ (v, k, m) $-FHS should
          have the property that all frequency channels used at any
          time slot $ t $ are uniformly distributed, that is we should
          have $ a_0 = a_1= \ldots = a_{m-1} $.
\end{description}
Recall that for a $ (\frac{1}{m}, \frac{1}{m}) $-adaptive jammer, what
happens at time $ t $ informs its next action at time $ t + 1 $,
therefore we also propose that:
\begin{description}
        \item[\namedlabel{m3}{\textbf{M3}}] For all FH sequences with
          frequency channel $ x_i $ at time slot $ t $, all frequency
          channels on the next time slot $ t+1 $ should be uniformly
          distributed. This forces a jammer to guess randomly at any
          time slot.
\end{description}
An FHS satisfying properties \ref{m2} and \ref{m3} would mean that a jammer
has no better chance of identifying an active channel at any time slot 
than randomly picking a channel.  Hence we would like our FHS to possess
these properties.  Properties \ref{m2} and \ref{m3} describe an 
orthogonal array:

\begin{defn}\cite{Hedayat99} 
A $k \times v$ array $A$ with entries from $\F$, 
$|\F|=m$, is said to be an
\textbf{orthogonal array} with $m$ levels, strength $ t' $, $ 1 \le t'
\le v $, and index $\lambda$ if every $k \times t'$ subarray of
$A$ contains each $t'$-tuple over $\F$ exactly $\lambda$ times as
a row.  We denoted this as $\OA_{\lambda} (k,v, m, t' )$.
\end{defn}

Clearly an $\OA_{\lambda} (k,v, m, t' )$ is also an 
$\OA_{\lambda m}(k,v, m, t'-1 )$ and $k = \lambda m^{t'}$.

Suppose we treat an $\OA_\lambda( k, v, m, t' ) $
as a $(v, k, m)$-FHS, $\Sn$. We want to know how long such an FHS can resist
a jammer, that is, we want to know how many time slots
a jammer would need to identify an active FH sequence.  

Consider first the situation where only one sequence is active in $\Sn$
and consider a $(\frac{1}{m}, \frac{1}{m})$-jammer in our FHS. At
$t=0$ (or indeed, at any time slot $t$), every channel in $\F$ appears $\lambda m^{t'-1}$
times, so the jammer has no better chance than randomly guessing a
channel $x$ to eavesdrop on. If $x$ is active we will say that the 
jammer is \textit{lucky}, otherwise we say that the jammer is \textit{unlucky}.

Now, if the jammer is lucky, that means that the active sequence must be
one of the $\lambda m^{t'-1}$ sequences with $x$ at $t=0$.  Thus the 
jammer would be able to reduce the 
search space for $t=1$ to $\lambda m^{t'-1}$, by a factor of $m$.
If the jammer is unlucky, then it will be able to remove 
from its search space the sequences with $x$ at time $t=0$, and on 
$t = 1 $ continues its search on the remaining sequences in $\Sn$. 
The search space would be reduced from $\lambda m^{t'}$ to 
$\lambda m^{t'} - \lambda m^{t'-1} = \lambda m^{t'}(m-1)/m $, 
a factor of $(m-1)/m$.   In summary,

\begin{lemma}
Let  $\Sn$ be a $(v, k, m)$-FHS which is also 
an $\OA_\lambda( k, v, m, t' )$.  Suppose only one sequence is
active in $\Sn$.  Then for a $(\frac{1}{m}, \frac{1}{m})$-jammer,
the size of the search space $\Sn_t^*$ at time $t$, $0\le t \le t'$, is given
by $ | \Sn^*_{t} | = \lambda (m-1)^B m^{t'-t} $ where $ B
$, $ 0 \le B \le t \le t' $, is the number of time slots in which a
jammer has been unlucky.
\end{lemma}

The jammer continues this action until either 
an active codeword is identified or the session ends.

One well-known class of orthogonal arrays is the
maximum distance separable (MDS) codes:
An $\OA_1 ( m^{t'}, v, m, t' ) $ is a $ (v, m^{t'}, m; v - t' + 1) $-MDS code.
Example \ref{eg:RScodes} gives an example of Reed-Solomon codes which are
MDS codes that are both orthogonal arrays and cover-free codes.  
In this case we may be more specific:

\begin{cor}
Let $\Sn$ be a $(v, k, m)$-FHS which is also an $\OA_1( k, v, m, t'
)$.  Suppose there is only one active sequence in $\Sn$.  Then a
$(\frac{1}{m}, \frac{1}{m})$-jammer which is lucky all the time will
be able to identify the active sequence in $t'$ times slots.  Otherwise
it will take more than $t'$ times slots to identify an active sequence.
\end{cor}

However, the analysis above does not apply when there are more than
one active sequence.  If the jammer eavesdrop on channel $x$, say, at time
slot $t=0$ and is unlucky, it can
rule out all sequences with $x$ at $t=0$, hence reducing the search
space for the next time slot.  If it is lucky, 
all it could conclude is that 
\textit{one} of the active sequences is one of the $\lambda m^{t'}$ sequences
that has $x$ at $t=0$.  It cannot rule out the possibility that there
are other active sequences that do not have $x$ at $t=0$.  Therefore it
cannot reduce the search space. However, this is related to the notion of
\emph{descendants} in fingerprinting codes.  We will explore this briefly
in the next section.

\subsection{Another model of FHS}

Let $\Sn$ be a $(v, k, m)$-FHS and suppose $w+1$ of the
$k$ sequences are in use.  Now, the jammer 
 wishes to identify one of these
active sequences.  A possible strategy for the jammer is:

\begin{enumerate}
\item Pick the first channel $e_0$ which has the highest number of occurences in $\Sn$.
\item For $t \ge 1$, Record $\eaves = (e_0, \ldots, e_{t-1})$.
  Suppose there are $t_1$ active channels at time $i_0, \ldots,
  i_{t_1-1}$, and $t_2$ inactive channels at time $j_0, \ldots,
  j_{t_2-1}$, $t_1+t_2=t$.
\item At time $t$, pick $e_t$ to eavesdrop on.
  \begin{enumerate}
  \item If $e_t$ is active, compile a collection of subsets of $\Sn$
    capable of giving rise to the active parts of $(e_0, \ldots,
    e_{t-1}, e_t)$ and attempt to identify an active sequence.
  \item If $e_t$ is inactive, compile a subset of $\Sn$, getting rid
    of sequences that cannot possibly be active given $(e_0, \ldots,
    e_{t-1}, e_t)$. 
  \end{enumerate}
\end{enumerate}

Our aim is to design $\Sn$ so that any $(w+1)$-subset is able to
withstand such an attack for as long as possible, that is, it takes
the jammer as long as possible to identify an active sequence.
One possible approach to this is to view $\Sn$ as a code and $\eaves$
as a ``descendant'' of subsets of $\Sn$.  An active
sequence should belong to some parent set.  We will introduce
some notation and terminology here to better discuss this 
(see \cite{codes} for an introduction to fingerprinting codes).

Let $\Sn = \{X^i = (x^i_0, \ldots, x^i_{v-1}) \; | \; x^i_j \in \F,
i=1, \ldots, k.\}$.  Treat $\Sn$ as a code and $X^i$ as codewords over $\F$.
Let $D = (d_0, \ldots, d_{v-1})$.  We say $D$ is a descendant of a
subset $\{X^0, \ldots, X^w\}$ if $d_i \in \{x^0_i,
\ldots,x^{w}_i\}$ for all $i = 0, \ldots, v-1$.  We call $\{X^0,
\ldots, X^{w}\}$ a parent set of $D$.  These terms are used to
define traceability codes and codes with the identifiable parent
property (IPP) - given a word $D$ we can identify at least
one codeword in $\Sn$ that gives rise to $D$.  This finds application
in traitor tracing (\cite{codes}).  

Here we would like the opposite - given a word $\eaves$ obtained by
the jammer we would like the jammer NOT to be able to identify a
parent.  In addition we would need to capture the partial and
sequential nature of how $\eaves$ is obtained.

\begin{defn}[Partial descendants and parent sets.]
Let $i_0, i_1, \ldots, i_{t_1-1} \in \{0, 1, \ldots, v-1\}$, $t_1 \le v$.  
We say that $D=(d_0, \ldots, d_{v-1})$ is an 
\emph{$(i_0, \ldots, i_{t_1-1})$-partial
descendant} of a set 
$\{X^0, \ldots, X^{w}\}$ if 
$d_{i_j} \in \{ x^0_{i_j}, \ldots, x^w_{i_j}\}$ for
$j \in \{0, \ldots, t_1-1\}$, ignoring all other positions.  

We  call $\{X^0, \ldots, X^{w}\}$ a \emph{$t_1$-partial parent set}.  
\end{defn}
  
We consider the properties we would like for $\Sn$ that would defend against
the jammer strategy.  Considering Step 3(a), we 
would like the following property:

\begin{description}
        \item[\namedlabel{m4}{\textbf{M4}}] 
The intersection
of all the $t_1$-partial parent set should be the empty set for as large a $t_1$
as possible (otherwise an active sequence is identified).  
This should hold for all positions $(i_0,
\ldots, i_{t_1-1})$.  
\end{description}

Considering Step 3(b), we have: 

\begin{description}
        \item[\namedlabel{m5}{\textbf{M5}}] 
For every inactive channel that the jammer eavesdrop on,  
if the jammer can discard some
codewords at this step then property \ref{m4} should
 be preserved after expurgation.
\end{description}

It is not clear at this stage how properties \ref{m4} and \ref{m5}
relates to properties of other types of fingerprinting codes.  We can,
however, make the following statement about the Reed-Solomon code of
Example \ref{eg:RScodes}:

\begin{thm}
Let $m$ be a prime power such that $m \ge v $. Let $\delta$ be
an integer, $0 < \delta \le v$. 
Let $\F $ be the finite field of cardinality $ m $ and let 
$ \alpha_1, \alpha_2, \ldots, \alpha_v \in \F $ be distinct. 
Define a length $ v $ Reed-Solomon code $ \Sn $ over $ \F $ by:
                
$$ \Sn = \left\lbrace \left( f(\alpha_1), f(\alpha_2), \ldots , f(\alpha_v)\right) : f \in \F[X] \mbox{ and deg }f < \delta \right\rbrace.$$

Treat $\Sn$ as a $(v, m^{\delta}, m)$-FHS.  Suppose there are $w+1$
active sequence.  If the jammer eavesdrop at $\delta (w+1)+1$ time slots and
is lucky in all of them, then the jammer can determine an active sequence.  
\end{thm}
\begin{proof}
For ease of notation we assume that the jammer was lucky at time
slots $t=0, \ldots, \delta (w+1)$ and the active channels are $\eaves = 
(e_0, \ldots, e_{\delta (w+1)})$.

Every $\delta$ positions determines a unique sequence, so by taking 
combinations of $\delta$ of the $e_i$s we can determine at least
$$ {\delta (w+1) \choose \delta}{\delta w \choose \delta}
\cdots {2 \delta \choose \delta}$$
parent sets, each containing $w+1$ sequences.

Since there are $w+1$ active sequences, at least one
active sequence would have contributed $\delta$ of the $e_i$s, it will
appear in at least one of the parent sets.  

Suppose all active sequences contributed the same number $\delta$
of $e_i$, then all the active sequences would appear somewhere in the
parent sets, and $e_{\delta (w+1)}$ would identify one of them.

Suppose there is one active sequence that contributes more than $\delta$
of the $e_i$ then this sequence would appear multiple times in some parent 
set and thus would be identified. 
\end{proof}

On the other hand, if we have a scheme that has properties \ref{m4},
\ref{m5}, that
means that too many codewords can give rise to $\eaves$.  This implies that
too many codewords have the same symbol at a position, which
contradicts the requirement for low correlation.  Hence there is a trade-off
between throughput and jamming resistance.  This is a subject that warrants
further research.  In the next section we consider FHSs that achieve
both high throughput and jammer-resistance, at the expense of computational
costs.

\section{A secure and efficient FHS}\label{Discuss}

Section \ref{jptycfc} demonstrates the limits of how secure an FH
scheme based on codes can be. While MDS codes give a guarantee of
throughput, they may not resist a jammer for very long.  One possible
solution would be to restart the FH scheme every $ \gamma v $ time
slots.  We therefore propose that in order to withstand the attack of
an adaptive jammer, some form of pseudorandomness must be
introduced. Indeed, the schemes in \cite{Nyirenda14} suggests that
pseudorandomness is a necessary component of an FH scheme secure
against an adaptive jammer. We include the description of one of the
schemes here to illustrate this.  This scheme is able to withstand an
adaptive jammer for the entire session, at the expense of additional
computational burden, and on the assumption of a secure pseudorandom
number generator.

The ``strongly resilient Latin square (sR-LS) scheme'' \cite{Nyirenda14}
is constucted using a Latin square:

Let $\F = \mathbb{Z}_v$, the set of integers modulo $v$. A Latin
square of order $ v $ defined over $\F$ is a $ v \times v $ array $L$ such 
that no element of $\F$ appears more than once in a row or in
any column of $L$. Suppose $ x \in \F $. Let $ L + x = [\beta_{ij}]_{v
  \times v} $ where $ \beta_{ij} = \alpha_{ij} + x \mod{v} $. Then $ L
+ x $ is also a Latin square.
        
The sR-LS scheme is constructed from a Latin square $ L =
[\alpha_{ij}]_{v \times v} $ of order $v $ over $\F$ 
as follows. 

Let $ K $ be a long term key
shared by all legitimate users. Let $g$ be a pseudorandom function
that takes as input $ K $, the session number $s$, and the current
time slot $t$, and outputs an element of $\F$. A slot key $ x_t $ is
generated at each time slot as $ x_t = g(K, s, t) $.
The FH sequences are given as:
        $$ X_i = ( \beta_{ij} ), $$ 
where $ \beta_{ij} = \alpha_{ij} + x_t \mod{v} $.
Note that the sR-LS scheme has $ v $ FH sequences used in a session, and a worst-case  $ (v - 1) $-throughput of $ 1 $. It can be viewed in two ways,
\begin{itemize}
        \item A collection of $ v $ $ (v,v,v;1) $-FHS which are MDS codes of minimum distance $ v $, where each FH scheme is used only once. 
        \item An MDS code with minimum distance $ 1 $, that is, a $ (v, v^v, v; 1) $-FHS. Only $ v $ sequences are used and these are determined by the pseudorandom number generator.
\end{itemize}

It can be seen that in this scheme the $ v $ frequency channels at each time slot are unique. Therefore we have the maximum achievable $ w $-throughput of $ 1 $ for any $ w $, $ 1 \le w < v $. Now, consider a $ (\frac{1}{m}, \frac{1}{m} )$-adaptive jammer. The jammer has no knowledge of $ K $, as it is shared by only the legitimate users. Further, a fresh pseudorandom number $ x_t $ is generated at each time slot. Therefore a $ ( \frac{1}{m}, \frac{1}{m} ) $-adaptive jammer cannot identify an active FH sequence being used at a time slot.  So we have an FH scheme that achieves maximum $ w $-throughput of $ 1 $ and can withstand a $ ( \frac{1}{m}, \frac{1}{m} ) $-adaptive jammer for the entire session, that is $ \gamma v = v $.

\section{Conclusion}\label{Conclude}
In this paper we have discussed FH schemes in the presence of both
legitimate sequences of the system as well as jamming sequences of an
adaptive jammer. We have provided a system model and jammer model
where the performance of the FH schemes can be analysed in the
presence of both mutual interfering FH sequences as well as jamming
sequences.
        
So, it is desirable to know the throughput of an FH scheme under these
circumstances. We explored using cover-free codes as FH schemes as
they give a lower bound on the worst-case throughput. Further we
considered mitigating strategies for cover-free codes to be used in
the presence of jamming. However, we showed that in the presence of our
adaptive jammer, the FH schemes based on cover-free codes do not
withstand the attack for long. With this analysis, and with the
example of the strongly resilient Latin square (sR-LS) scheme proposed
in \cite{Nyirenda14} we conclude that pseudorandomness is a neccesity
in providing jamming resistance of FH sequences.

\end{document}